\documentstyle[12pt]{article}
\def\beq{\begin{equation}}
\def\eeq{\end{equation}}
\def\bea{\begin{eqnarray}}
\def\eea{\end{eqnarray}}
\def\bq{\begin{quote}}
\def\eq{\end{quote}}

\makeatletter
\def\vereq#1#2{\lower3pt\vbox{\baselineskip1.5pt \lineskip1.5pt
\ialign{$\m@th#1\hfill##\hfil$\crcr#2\crcr\sim\crcr}}}
\makeatother

\begin{document}

\begin{titlepage}
\begin{center}
\today     \hfill    SLAC-PUB-7801\\
~{} \hfill SU-ITP-98/28\\
~{} \hfill CPTH--S608.0498\\
~{} \hfill IC/98/39\\
~{} \hfill hep-ph/9804398\\
 
\vskip .1in

{\large \bf New Dimensions at a Millimeter to a Fermi\\
and Superstrings at a TeV}

\vskip 0.1in

Ignatios Antoniadis$^a$, Nima Arkani-Hamed$^b$, Savas Dimopoulos$^c$,
Gia Dvali$^d$

\vskip .05in
{\em $^a$ Centre de Physique Theorique, Ecole Polytechnique,\\
F-91128 Palaiseau, France}
\vskip .1truecm
{\em $^b$ Stanford Linear Accelerator Center, Stanford University,\\
Stanford, California 94309, USA}
\vskip 0.1truecm
{\em $^c$ Physics Department, Stanford University,\\
Stanford, California 94309, USA} 
\vskip 0.1truecm
{\em $^d$ ICTP, Trieste, 34100, Italy}
\end{center}

\begin{abstract}
Recently, a new framework for solving the 
hierarchy problem has been proposed which does not
rely on low energy supersymmetry or technicolor.
The gravitational and gauge interactions unite at the
electroweak scale, and the observed 
weakness of gravity at long distances is due the existence of 
large new spatial dimensions. 
In this letter, we show that this framework can be embedded in 
string theory.  
These models have a perturbative description in the
context of type I string theory. The
gravitational sector consists of closed strings propagating in the
higher-dimensional bulk, while ordinary matter consists of open
strings living on D3-branes. 
This scenario raises the exciting possibility that the LHC and NLC 
will experimentally study ordinary aspects of string physics 
such as the production of narrow Regge-excitations of all 
standard model particles, as well more exotic phenomena 
involving strong
gravity such as the production of black holes. 
The new dimensions can
be probed by events with large missing energy carried off by gravitons 
escaping into the bulk.
We finally discuss some important issues of model
building, such as proton stability, gauge coupling unification and
supersymmetry breaking.

\end{abstract}

\end{titlepage}

\def\simlt{\stackrel{<}{{}_\sim}}
\def\simgt{\stackrel{>}{{}_\sim}}
\newcommand{\cm}{Commun.\ Math.\ Phys.~}
\newcommand{\prl}{Phys.\ Rev.\ Lett.~}
\newcommand{\pr}{Phys.\ Rev.\ D~}
\newcommand{\pl}{Phys.\ Lett.\ B~}
\newcommand{\ibar}{\bar{\imath}}
\newcommand{\jbar}{\bar{\jmath}}
\newcommand{\np}{Nucl.\ Phys.\ B~}
\newcommand{\be}{\begin{equation}}
\newcommand{\en}{\end{equation}}
\newcommand{\ba}{\begin{eqnarray}}
\newcommand{\ea}{\end{eqnarray}}
\newcommand{\aG}{\alpha_G}

\section{Introduction} 
In a recent paper \cite{ADD1}, a general framework for solving the hierarchy
problem was proposed not relying on low-energy supersymmetry or technicolor.
The hierarchy problem is solved by nullification:
in this scenario, gravity 
becomes unified with the gauge interactions at the weak scale
and 
there is no large disparity 
between the size of different short distance scales in the theory.
As argued in \cite{ADD1}, the observed weakness of gravity is then due to the
existence of new spatial dimensions much larger than the weak scale,
perhaps as large as a millimeter for the case of two extra dimensions.
The success of the Standard Model (SM) 
 then implies that, while gravity 
is free to propagate in the bulk of the extra dimensions, the SM fields must
be localised to a 3 spatial dimensional wall at energies beneath the weak scale.
While field-theoretic mechanisms for 
localising the SM fields on a topological defect were suggested,
the nature of the theory of gravity above the weak scale 
was left unspecified in the 
general framework of \cite{ADD1}.

In this letter, we show that the  
above scenario can be embedded within string theory, which at present offers
the only hope for a consistent theory of gravity. 
The traditional line of thought has been 
that string theory becomes relevant only
at very short distances of order of the Planck length $\sim 10^{-33}$ cm.
However various arguments involving unification,
supersymmetry breaking or the gauge hierarchy, suggest that it may
be relevant at even larger distances. 
For instance, compatibility
of string unification with gauge coupling unification 
within the minimal supersymmetric standard model \cite{DG}
implies that
the string (or M-theory) scale should be of the order 
$M_{\rm GUT}\sim 10^{16}$ GeV, while additional
dimensions would show up at even lower energies $\simlt 10^{15}$ GeV
\cite{witten,aq}. As another example, low energy supersymmetry breaking 
within perturbative string theory implies the existence of a large internal
dimension whose size determines the breaking scale \cite{a}. 
The possibility that the string
scale is close to the electroweak scale was mentioned in, for example
\cite{b,Lykken,aq}. This is certainly a requirement for a string
realization of the scenario proposed in \cite{ADD1}. 

In this work we show that
the only perturbative string theory with weak scale string tension must be 
a type I theory of open and closed strings with
(a) new dimensions much larger 
than the weak scale ranging from a fermi to a millimeter (b)
an $O(1)$ string coupling and (c) SM fields identified with 
open strings localised on a 3-brane. Aside from 
providing a specific realization of 
the considerations of \cite{ADD1}, this construction 
has the immediate advantage that the localisation 
of non-gravitational fields to a three dimensional submanifold 
is automatic and natural.
Moreover, our explicit
realization will allow us to address a certain
number of important theoretical questions
arising from this idea, while simultaneously it
offers a calculable framework for studying its
phenomenological implications.

\section{String embedding} 
Any perturbative description of string theory
has two fundamental parameters: the string scale
$M$ and a dimensionless coupling $\lambda$ which
controls the loop expansion.\footnote{Actually
$\lambda$ corresponds to the vacuum expectation
value of a dynamical scalar field.} Upon
compactification to $D=4$ dimensions, these
parameters can be expressed in terms of the 4D
Planck mass $M_p$, the gauge coupling $\alpha_G$
at the string scale and the compactification
volume $(2\pi)^6V$ of the internal
six-dimensional manifold. Imposing the string
scale to be at a TeV one can in
principle solve for $V$ and $\lambda$ and, then,
trust the solution if $\lambda<1$. However in
the weakly coupled heterotic theory, $V$ and
$\lambda$ drop from one of the two relations and
one obtains a prediction for the string scale
$M_H$: $M_H=({\alpha_G\over 8})^{1/2}M_p\simeq
10^{18} {\rm GeV}$\footnote{In fact $\alpha_G$ should be replaced
by $k\alpha_G$, where $k$ is the integer Kac-Moody level.}.

The situation changes drastically in the
strongly coupled heterotic theory. The latter is
described, in the $E_8\times E_8$ case, by the
eleven-dimensional M-theory compactified on a
line segment of length $\pi R_{11}$ \cite{HW}. 
One may now
try to identify the M-theory scale $M_{11}$,
which determines the Newton constant of the 11D
supergravity, with the electroweak scale.
Unfortunately, then, the length of the line
segment turns out to be unacceptably large:
$R_{11}={\alpha_G\over 2}{M_p^2\over
M_{11}^3}\simeq 10^8{\rm km}$!

It only remains to consider the type I theory of open and closed
strings which also decribes the strongly coupled
heterotic $SO(32)$ string \cite{pw}. 
When compactified down to four dimensions,
the gravitional and gauge kinetic terms 
of the resulting effective four-dimensional theory are, 
in a self-explanatory notation:
\be
S=\int d^4x\sqrt{-g}\left(
{1\over\lambda_I^2}V M_I^8{\cal R}
+{1\over\lambda_I}V M_I^6F^2\right)\, ,
\label{SD}
\en
where we ommitted numerical factors. Identifying the coefficient of 
${\cal R}$ with $M_{p}^2$ and that of $F^2$ with $1/g^2$ yields the 
relations   
\be
V^{-1}={\alpha_G^2\over 2}M_p^2 M_I^4\quad\quad
;\quad\quad
\lambda_I={8\over\aG}\left({M_I\over M_p}\right)^2\, ,
\label{typeI}
\en
where in this equation the relevant numerical factors have
been included. 
Taking the type I string scale $M_I$ to be at
the TeV, one finds a compactification scale much
larger, while the string coupling is
infinitesimaly small. Choosing for instance $n$
internal dimensions to have a common radius
$R_I$ and the remaining $6-n$ of the string
size, one obtains:
\be
R_I^{-1}=\left({\aG^2\over
2}\right)^{1/n}M_p^{2/n}M_I^{1-2/n}
\qquad n=1,\dots,6\, .
\label{RI}
\en
It follows that for $\aG\simeq .1$ the value of
the compactification scale varies from $10^{33}$
GeV, $10^{18}$ GeV, up to $10^8$ GeV for $n=1$,
2, or 6 large internal dimensions.

One may naively think that such a large
compactification scale is irrelevant for low
energy physics. However this is not true in
string theory due to the presence of winding
states in the closed string (gravitational)
sector, whose masses are quantized linearly with
the radius in string units and are therefore
very light. In fact physics is equivalent as if
there was a radius $R=1/(R_IM_I^2)$ which is
much larger than the string length and the roles
of windings and Kaluza-Klein (KK) momenta are
interchanged. On the other hand, in this T-dual
theory, open string states which give rise to
ordinary non gravitational matter live on 
D3-branes, since they have only winding modes,
identical with the heavy KK modes (\ref{RI}) of
the initial theory.

By performing a T-duality to all six compact
dimensions, $R\to 1/(RM^2)$ and
$\lambda\to\lambda/(RM)$, the relations
(\ref{typeI}-\ref{RI}) become:
\ba
V^{-1} &=& {2\over\alpha_G^2}M_p^{-2}
M_I^8\quad\quad
;\quad\quad\lambda=4\aG
\nonumber\\
R^{-1} &=& \left( {2\over\aG^2} \right)^{1/n}
M_I\left({M_I\over M_p}\right)^{2/n} <<M_I\, .
\label{Tdual}
\ea
As a result, the coupling constant of this dual
theory is the 4D gauge coupling while the value
of the compactification scale varies from
$10^{-18}$ eV, $10^{-3}$ eV, up to $10$ MeV for
$n=1$, 2, or 6 large internal dimensions.
Obviously the case of $n=1$ is experimentally
excluded, $n=2$ corresponds to two dimensions in
the range of 100 microns, while $n=6$ to six
dimensions in the range of .1 fermi.

This setup gives an explicit string
realization of the proposal of \cite{ADD1}.
In fact,
from eq.~(\ref{Tdual}) we have $M_I\sim
(\aG^2M_p^2
R^{-n})^{1/(n+2)}\equiv\aG^{2/(n+2)}M_{p(4+n)}$,
where $M_{p(4+n)}$ is the Planck mass in the
($4+n$) higher dimensional theory. Actually in
string theory the presence of the gauge coupling
in the above relations lead to a string scale a
bit lower that the ($4+n$)-dimensional Planck
mass. Taking $M_I$ at the TeV, this leads to
somewhat lower values for $R$ when $n$ is small.

An immediate advantage of the string
construction is that matter is automatically
localized on a 3-brane. The latter can be
thought as a thin wall limit of the effective
field theory solution. In fact, an important
difference here is that all string states living
on the brane are delocalized in the extra large dimensions only at energies
much higher than the string scale, $E\sim
RM_I^2=1/R_I$, due to the effect of string
winding modes. Above the compactification scale
(\ref{RI}) of the dual theory, $n$ extra
dimensions open up for matter, as well.
Of course this statement should be understood
under the assumption than one can naively
extrapolate at energies above the string scale,
excluding non perturbative effects and the possibility of phase transitions,
especially if the theory is not supersymmetric.

It is interesting that, in compactifying 
to four dimensions, the only way of making the 
string scale much lower than the (4 dimensional) Planck 
scale is to have large volume compactifications, with $O(1)$
string coupling. One may wonder in what cases the new dimensions
can be kept at the string scale, with an infinitesinal string coupling
accounting for the discrepancy between the Planck and weak scales. 
For the general case of compactifying on a $10 - D$ dimensional 
manifold, the relations generalizing eqn.(\ref{typeI}) read (again
omitting numerical factors)
\be
V_{10-D}^{-1}=\alpha_D^2
M_{p(D)}^{6-D}M_I^4\quad\quad ;\quad\quad
\lambda_I={1\over\alpha_D}\left({M_I\over M_{p(D)}}\right)^2\, ,
\label{typeID}
\en 
It follows that precisely at $D=6$ the internal
volume becomes of order of the string scale,
which can be much smaller than the 6D Planck
mass by tuning only the string coupling, while
keeping the dimensionless gauge coupling to be
of order unity. This is consistent with the
six-dimensional examples considered in \cite{Lykken}.
For $D>6$ the compactification scale is smaller than
$M_I$, while for $D<6$ it becomes bigger and 
(as we did in the case $D=4$) one
has to go to the T-dual theory in order to
describe low energy physics with an effective
field theory.

\section{Phenomenology}
\subsection{Accelerator signals and constraints}

There are two distinct classes of novel
phenomena that occur at a TeV in our
framework\footnote{These, together with
laboratory and astrophysical constraints on the more general framework 
proposed in \cite{ADD1} are
discussed in detail in ref.\cite{ADD2}}:\\
(1)Production of Regge-excitations\\
(2)Emission of $(4+n)$-dimensional gravitons
into the extra dimensions.\\

The existence of Regge-excitations (RE) for all
the elementary particles of the standard model
is a consequence of having a string theory at a
TeV and does not per-se reveal the existence of
extra dimensions. In contrast to the RE of
ordinary QCD, these states are expected to be
relatively weakly coupled and narrow since the
string coupling constant is not too large. The
ratio of their width to their mass is of order
of $\Gamma/m \sim \lambda^2 \sim$ a few per thousand, 
so they are relatively narrow resonances with well defined
mass. The Regge-excitations of the gluon could
be produced in gluon-gluon collisions at LHC,
showing up as a series of narrow peaks
in the gluon-gluon cross section as a function
of the gluon pairs' invariant mass. 
The corresponding amplitude is proportional to:
\begin{equation}
A(s,t) \sim {\Gamma(-s/M_I^2)\Gamma(-t/M_I^2)\over\Gamma(1-(s+t)/M_I^2)}\, ,
\label{A}
\end{equation}
exhibiting a series of poles corresponding to the RE mass positions. 

Next we come to graviton emission into the extra
dimensions \cite{ADD2}. The inclusive cross
section for single graviton emission is proportional to 
\begin{equation}
\sigma(E) \sim \frac{E^n}{M_I^{n+2}} \times (n+2)^2 
\label{cross}
\end{equation} 
where the $M_I$ dependence is uniquely fixed
by the normalization of the graviton fields in the higher dimension theory
$g_{AB} = \eta_{AB} + h_{AB}/\sqrt{M_I^{n+2}}$.
A $(4+n)$-dimensional graviton contains, in
addition to the normal 4-D graviton,
graviphotons, Brans-Dicke scalars as well as an
antisymmetric tensor. In the present case
all of these particles can be
emitted into the extra dimensions with equal
amplitudes. This leads to the multiplicity
factor proportional to $(n+2)^2$ of eq.(\ref{cross}) and
enhances the total graviton emission rate.

The cross section (\ref{cross}) is negligible for energies much
below $M_I$ but rises abruptly at energies of
order $M_I$, leading to an abundance of events with
lots of missing energy, carried by gravitons
into the extra dimensions. One way to look for
these in hadron colliders is to search for
processes with jets+missing energy. Another
manifestation of graviton emission into the
extra dimensions is that it leads to phenomena
analogous to those caused by brehmstrahlung. 
{}For example graviton emission,
or ``gravistrahlung'', depletes the beam energy
just before a collision takes place. This should
be taken into account in the resonant production
of narrow states, such as the Regge-excitations 
of the gluon in gluon-gluon collisions and of
other ordinary particles. 

Graviton emission will
be very important at energies above $M_I$ where it
is analogous to Hawking radiation from an
excited brane and can be computed using the
technology developed, for example, in
ref.\cite{Klebanov}:
\begin{equation}
\sigma(E) \sim (n+2)^2 \frac{E^n}{M_I^{n+2}}
{\Gamma(1-2E^2/M_I^2)^2\over\Gamma(1-E^2/M_I^2)^4}\, .
\label{cross1}
\end{equation} 
The emmission rate exhibits a sequence of poles associated to the
production of RE resonances, as well as a sequence of zeros
indicating that the corresponding states are forbidden to propagate
in this process \cite{Klebanov}. Moreover, it decays exponentially
at large energies due to the well known
ultraviolet softness of string theory.

There is a qualitative difference between the
present D-brane construction and the 
``thick wall" version of the scenario 
proposed in \cite{ADD1}. In the
latter the binding energy of particles to the
walls is typically of order of the
weak scale and therefore the particles of
the Standard Model can be emitted into the bulk
in collisions at TeV energies. This leads to
qualitatively different, and even more dramatic,
missing energy signatures than those of the
D-brane case where only the gravitons migrate in
the bulk. In our case, however, 
the SM particles are open strings stuck on the brane. 
They do possess winding modes that feel the bulk, but their mass
is very large $\sim RM_I^2$, which ranges from $10^8$ GeV
for $n=6$ to $10^{19}$ GeV for $n=2$. These winding modes will then be
irrelevant to weak scale physics.

Finally a comment on phenomenological
constraints on these theories. There are several
which are discussed in ref.\cite{ADD2}. The most
obvious, but not the most important, comes from the
compositeness bounds on the scale 
of suppression of higher dimension operators, which are
at most at 3 TeV for flavor-conserving operators. Such effects
are induced by the exchange of the
Regge-excitations  as well as the graviton. For
the former the most dangerous effect comes from
the exchange of the RE of the photon between two
electrons. It is safely small as long as the RE
of the photon are heavier than 300 GeV. The
exchange of a $(4+n)$-dimensional graviton between
two electrons induces a higher dimension operator of the 
form:
\begin{eqnarray}
{\cal O} & \sim &  \left(\frac{E^n}{M_I^{n+2}}\right) \times 
\left(\frac{M_I}{E}\right)^{n-2} \times (n+2)^2 (\bar{\psi} \psi)^2
\nonumber\\
&\sim& (n+2)^2 \frac{E^2}{M_I^4} (\bar{\psi} \psi)^2,
\end{eqnarray} 
where in the first line of the above equation, the 
first factor has the naive $M_I$ dependence following from the normalisation
of $h_{AB}$ and the second results from 
the sum over the heavy KK excitations 
of the graviton with mass greater than $E$ which is UV divergent for 
$n \geq 2$ \cite{ignatB}.  
Since the largest energy, where the 4-electron
vertex is studied accurately, is $\sim 100$ GeV this is
safely small provided the string scale $M_I$ is
larger than $\sim$ 1 TeV.

\subsection{Proton stability}
Every extension of the SM invoking new physics at the 
electroweak scale must address the issue 
of the stability of the proton and the abscence of 
large flavor-violations. An arbitrary effective Lagrangian with 
all higher dimension operators suppressed by powers of the weak scale 
is grossly excluded: assuming $O(1)$ coefficients, 
the standard dimension 6 operators
giving proton decay must be suppressed by at least the GUT scale to be safe,
while operators contributing to $\Delta m_K$ and $\epsilon_K$ must be suppressed
by $\sim 10^3$ and $10^4$ TeV respectively. Clearly some mechanism is required 
to adequately suppress these operators. Since the size of flavor-changing
operators is intimately linked to the origin of flavor, the hope is that the 
same physics which suppresses the light generation Yukawa couplings also
adequately suppresses the FCNC operators. We will therefore 
not pursue the flavor-changing issue here, 
and focus on the far more serious problem of proton decay. 

Adequately suppressing baryon number violating operators is 
somehwat easier when the theory above the weak scale is a known field theory
as in the case of the MSSM, since only the dangerous dimension 4 operators must
crucially be forbidden, and this can be arranged fairly simply (e.g. by imposing
$R$ parity). One certainly does not have to impose a symmetry forbidding proton
decay altogether. In our case, however, the theory above the weak scale is 
an unknown string theory, and it is not clear if there is some simple
mechansim analogous to imposing $R$ parity 
which adequately suppresses proton decay without
forbidding it. Without knowing the structure of the full theory,  
it seems safer to look for symmetries that can be imposed on the 
low energy theory which completely shut off proton decay. 
Of course, one can imagine
that $U(1)_B$ is an exact global symmetry of the theory 
beneath the string scale, but 
standard lore suggests that all symmetries in string theory are local, and we
will limit ourselves to this possibility. We list a few possibilities below.

A simple possibility \cite{HitChris} is to add a fourth generation whose quarks
are assigned baryon number $1$ instead of $-1/3$; $U(1)_B$ is then the 
diag(1/3,1/3,1/3,-1) generator of the flavor $SU(4)$ acting on the quarks.
It is easy to see that with this assignement, $U(1)_B$ is anomaly free.
Of course, we don't want to gauge the continuous $U(1)_B$ since no massless
gauge boson coupled to baryon number has been observed. We can however gauge 
any discrete $Z_q$ subgroup of this $U(1)_B$, arising for 
instance if $U(1)_B$ is broken by a scalar field of charge $q$, implying 
baryon number is conserved modulo $q$. Proton decay is completely forbidden as
long as $q \neq 1$, and other baryon number violating operators
can be enormously suppressed.
For instance, in \cite{HitChris}, $U(1)_B$ is broken by a field of charge 
$4/3$ in order to allow ordinary Yukawa couplings of the 4th to the rest of
the generations, so $B$ is conserved mod $4/3$. Since all physical states have
integer baryon number this means that in physical processes $B$ 
is conserved mod $4$, and lowest dimension operator invariant under 
$Z_{4/3}$ but violating $U(1)_B$ is $(QQ)^4 D_c^{\dagger 4}$ has dimension 18, 
and is certainly safe even when 
suppressed only by the weak scale.

There are other possibilities which do not require the addition of new chiral
matter to the SM. For instance, Ref. \cite{IbanezRoss} gives an 
anomaly-free discrete ``baryon triality" acting as $(1,g^2,g,g^2,g^2)$ on
$(Q,U^c,D^c,L,E^c)$, where $g^3=1$. 
It is easy to check that 
all operators invariant under the SM and this triality 
violate baryon number by a multiple of 3 units: the
lowest dimension operator violating $B$ by 3 units is 
18 dimensional and certainly safe. 

\subsection{Gauge coupling unification}
Since the string scale is close to the weak scale in our scenario, the usual 
logarithmic evolution of the gauge couplings can not be extrapolated to high 
scales and the standard picture of gauge coupling unification is lost. 
On the other hand, there are interesting new possibilities for gauge coupling
unification at the weak scale, provided that the $SM$ fields stuck to the 
brane can propagate in $k$ spatial dimensions with a size $r$ somewhat larger than
the string scale. The effective theory above energies $r^{-1}$ but beneath 
the string scale is a higher dimensional field theory, and since gauge couplings
are dimensionful in higher dimensions, they run with a power of energy, rather
than logaritmically as in four dimensions. 
This can also be understood from the four dimensional viewpoint due
to the contribution to the running of the towers of KK excitations: 
at energies $E>1/r$, we have (at 1-loop)
\begin{equation}
E \frac{d g}{dE} = \frac{b + n(E) b_{KK}}{8 \pi^2}
\end{equation}
where $n(E) \sim (Er)^k$ is the number of KK modes which are lighter than
energy $E$ and $b_{KK}$ is the contribution to the beta function from 
one KK level.
Depending on the quantum numbers of the KK towers,
the couplings can evolve very quickly above the scale $
1/r$ to a unified value at the string scale \cite{Dienes}.

\subsection{SUSY breaking} 
In our scenario, there is no longer any need for a ``hierarchical" breaking of
SUSY, since the string scale is already close to the scale where SUSY should be
broken, a feature we find attractive relative to the standard scenario.
On the other hand, some of the radii of compactification must be very large
compared to the string scale, and while this may have a cosmological explanation,
we wish to leave open the possibility that the moduli corresponding to the 
size of these dimensions are stabilised at large values. Since the
potential for the moduli vanish in the SUSY limit, it seems natural to keep the 
size of SUSY breaking for the modulus very much smaller than the TeV scale.
This can be done if SUSY is primordially broken only on the brane,
with SUSY breaking transmitted to the bulk by gravitational effects.
If SUSY is broken maximally on the brane (i.e. 
$\rho_{brane} \sim M_{st}^4$), and assuming
that the transmission to the bulk fields occurs at tree level
(therefore proportional to $G_{N(4+n)}$), we can estimate an upper limit
the SUSY breaking soft mass for the modulus by dimensional analysis
\begin{equation}
m^2_{modulus} \sim \frac{G_{N(4+n)} \rho_{brane}}{R^n} \sim 
\frac{\rho_{brane}}
{M_{p}} \sim (1 \mbox{mm})^{-2}.
\end{equation}
The spectrum in the bulk is very nearly (at least $N=2$)
supersymmetric, with 
bose-fermi splittings of at most $\sim (1$mm)$^{-1} \sim 10^{-4}$eV.
It is reassuring that the modulus mass is at most the inverse size of the 
compact dimensions for $n=2$ or smaller for $n>2$.

The presence of such a light modulus would generate gravitational forces in the
millimeter range which could be explored experimentally \cite{forces}. Moreover,
in the case of $n=2$ large dimensions, one should take into account all the 
components of the six-dimensional graviton and of other fields, giving rise to
additional scalars and graviphotons. Since the latter contribute to a repulsive
force, the gravitational interactions might be drastically modified at 
(sub)millimeter distances. In particular, if the bulk had $N=4$ supersymmetry, 
one would have an exact cancellation of gravitational forces at short distances!

\section{Conclusions}

As will be discussed in greater detail
in \cite{ADD2}, the scenario we propose is not experimentally excluded
by any lab or astrophysical constraint we are aware of. We briefly discussed
some issues of model building such as stabilising 
the proton through discrete gauge symmetries, 
SUSY breaking and gauge coupling
unification at the weak scale, but it is clear that much work remains to be
done to construct a completely realistic model. 
In particular, the most pressing theoretical issue is to understand the 
origin of the large size of the extra dimensions, ranging from 5 to 15 orders
of magnitude larger than the string scale. 
Given the prospect 
of studying quantum gravity at the LHC and NLC, we feel that future 
model-building within this framework is well motivated and exciting.

\section{Acknowledgements}
We wish to thank Petr Horava for valuable correspondence.
I.A. and G.D. would like to thank the Institute of 
Theoretical Physics at Stanford for their hospitality. 
NAH is supported by the Department of Energy under contract 
DE-AC03-76SF00515. SD is supported by NSF grant PHY-9219345-004.
IA and GD are supported partially by the European Community under the
TMR contract ERBFMRX-CT96-0090.

\def\pl#1#2#3{{\it Phys. Lett. }{\bf B#1~}(19#2)~#3}
\def\zp#1#2#3{{\it Z. Phys. }{\bf C#1~}(19#2)~#3}
\def\prl#1#2#3{{\it Phys. Rev. Lett. }{\bf #1~}(19#2)~#3}
\def\rmp#1#2#3{{\it Rev. Mod. Phys. }{\bf #1~}(19#2)~#3}
\def\prep#1#2#3{{\it Phys. Rep. }{\bf #1~}(19#2)~#3}
\def\pr#1#2#3{{\it Phys. Rev. }{\bf D#1~}(19#2)~#3}
\def\np#1#2#3{{\it Nucl. Phys. }{\bf B#1~}(19#2)~#3}
\def\mpl#1#2#3{{\it Mod. Phys. Lett. }{\bf #1~}(19#2)~#3}
\def\arnps#1#2#3{{\it Annu. Rev. Nucl. Part. Sci. }{\bf #1~}(19#2)~#3}
\def\sjnp#1#2#3{{\it Sov. J. Nucl. Phys. }{\bf #1~}(19#2)~#3}
\def\jetp#1#2#3{{\it JETP Lett. }{\bf #1~}(19#2)~#3}
\def\app#1#2#3{{\it Acta Phys. Polon. }{\bf #1~}(19#2)~#3}
\def\rnc#1#2#3{{\it Riv. Nuovo Cim. }{\bf #1~}(19#2)~#3}
\def\ap#1#2#3{{\it Ann. Phys. }{\bf #1~}(19#2)~#3}
\def\ptp#1#2#3{{\it Prog. Theor. Phys. }{\bf #1~}(19#2)~#3}

\end{document}